# Award rate inequities in biomedical research

Alessandra Zimmermann[1], Richard Klavans[2], Heather Offhaus[3,¶], Teri A. Grieb[3,¶], and Caleb Smith[3,*]

[1] Proposal Analytics, Inc., Wayne, PA, USA
[2] SciTech Strategies, Wayne, PA, USA
[3] Office of Research, University of Michigan Medical School, Ann Arbor, MI, USA
[*] Corresponding author, Email: calebs@umich.edu (CS)

# Abstract

## Purpose

The analysis of existing institutional research proposal databases can provide novel insights into science funding parity. The purpose of this study was to analyze the relationship between race/ethnicity and extramural research proposal and award rates across a medical school faculty and to determine whether there was evidence that researchers changed their submission strategies because of differential inequities across submission categories.

## Method

The authors performed an analysis of 14,263 biomedical research proposals with proposed start dates between 2010-2022 from the University of Michigan Medical School, measuring the proposal submission and award rates for each racial/ethnic group across 4 possible submission categories (R01 & Equivalent programs, other federal, industry, and non-profit).

## Results

Researchers from each self-identified racial/ethnic group (Asian, Black/African American, Hispanic/Latino) pursued a different proposal submission strategy than the majority group (White). The authors found that Black/African American researchers experienced negative award rate differentials across all submission categories, which resulted in the lowest R01 & Equivalent and Other Federal submission rates of any racial/ethnic group and the highest submission rate to non-profit sources. The authors did not find support for the hypothesis that researchers changed submission strategies in response to award rate inequalities across submission categories.

## Conclusions

Biomedical researchers from different racial/ethnic groups follow markedly different proposal submission strategies within the University of Michigan Medical School. There is also a clear relationship between race/ethnicity and rates of proposal award. Black/African American and Asian researchers appear disadvantaged across all submission categories relative to White researchers. This study can be easily replicated by other academic research institutions, revealing opportunities for positive intervention.

# Introduction

The scope and scale of academic research pursued within the United States today is not possible without federal funding. The National Science Foundation's annual Higher Education Research and Development Survey reported total higher education R&D expenditures of $86.4 billion for federal fiscal year 2020 with more than 46 of those expended billions originating with the federal government [1]. Focusing more specifically on academic biomedical research, the National Institutes of Health (NIH) alone awarded research grants in excess of $29 billion during the same fiscal year [2]. It is not surprising then that when the first major study documenting racial disparity in research funding was published in 2011, the authors focused on the NIH [3]. Since this original study, others have added to the national discussion using NIH proposal data [4-8].

The NIH is the largest and most influential funder of academic biomedical research within the United States. NIH research awards, particularly R01 and equivalent funding mechanisms, play an important role in promotion and tenure decisions as well as in the stability of research programs. The size and duration of these awards allow researchers to spend more of their effort on research and less on grant-writing. Academic institutions particularly value NIH award funding as it generally comes with higher indirect cost rates than non-federal funding and is used in external rankings of Medical School research programs. For instance, when evaluating Medical Schools for their "Best Medical Schools for Research" rankings, US News & World Reports assigns a weight of 0.4 out of 1.0 to "Research Activity". This metric is based solely on federal funding [9], the large majority of which comprises NIH grants. While US News & World Reports began publishing a campus ethnic diversity index in 2020 [10] it is not incorporated into the "Best Medical Schools for Research" calculation and does not currently influence these rankings. As Ginther has shown [3,4] and researchers within the NIH have recently acknowledged and expanded upon [5,6], there remains a significant gap between NIH funding rates of white principal investigators (PIs) and Black/African American (B/AA) PIs. As such, there is an inherent tension between medical school rankings and faculty diversity.

The need for increased diversity in academic medicine and the biomedical workforce has been well substantiated. While many academic institutions advocate for increased equity among faculty sub-demographics, there remains a dearth of quantitative information supporting specific interventions. Understandably, academic administrators and faculty leaders are hesitant to implement reforms that could unintentionally compound existing inequities or even create new systemic problems. The issues are considered too important to get wrong. This sense of policy paralysis, in turn, can lead institutions into a cycle of calls for action followed by a series of educational events that, while important, are insufficient on their own to effect policy change.

One practical step forward is for academic administrators to utilize the proposal databases that have already been created at their institutions to understand and evaluate diversity in academic research. Given the importance of extramural grant funding to academic research and the subsequent emphasis placed upon grant funding in everything from individual faculty promotion and tenure review to departmental evaluation, it is reasonable to begin any study of potential research inequity with an analysis of research grant proposal submissions and awards. This will not only increase the collective knowledge about how our research funding system works in all its complexity, but it will also, perhaps most importantly, represent a starting point for changing faculty development and research support procedures, performance metrics and their interpretation, as well as promotion and tenure review systems.

Dr. James Hildreth noted that at historically black colleges, there is an "intimidation factor" about submitting to the NIH and a presumption that proposals will not be evaluated fairly [11]. This anecdote, along with the reported racial/ethnic differences in award rates by the NIH, led us to investigate the relationship between race/ethnicity and submission behaviors for principal investigators at the University of Michigan Medical School (UMMS). We hypothesized that principal investigators adjust their proposal submission strategies to compensate for Award Rate Inequality (ARI), established or perceived from lived experience, across funding sources.

# Materials and methods
## Data sources and features

These analyses are based on the University of Michigan's electronic proposal management system, a proposal database which stores information about each extramural research proposal submitted by University of Michigan faculty. This system is used for the routing, institutional approval, and submission of all extramural research projects at the University. Among the attributes of the research proposals tracked within this system are unique identifiers for project personnel, which allow us to annotate each proposal record with self-reported demographics, such as the race and ethnicity of the principal investigator.

Ethical approval for this study was granted by the University of Michigan Medical School Institutional Review Board (Study ID: HUM00200080, Federalwide Assurance: FWA0000496). The review board waived the requirement for informed consent and determined that the study was exempt from ongoing review following federal exemption category 4(ii) at 45 CFR 46.104(d).

Our sample comprises 14,263 unique research proposals with proposed start dates after 01/01/2010. We exclude subcontracts, clinical trials, and proposals for extramural funding for activities other than research from our observations. Further, we limit our data to only those proposals awarded or turned down by the sponsor. Proposals withdrawn by the PI as well as those proposals with pending sponsor decisions at the time of our query are excluded. Table 1 lists the distribution of the seven racial and ethnic categories selected by PIs of the remaining proposals within this study. In the event that more than one PI is listed on a proposal, only demographic details of the PI designated as the "Contact PI"—the PI to whom questions about the proposal should be addressed—are used for this study. Three categories are excluded from further analysis due to insufficient sample size or because race/ethnicity cannot be determined. Over 98% of all principal investigators within the original sample self-identify with one of the four remaining racial/ethnic groups.

**Table 1. UMMS Racial/Ethnic Groups (2010-2022).**

| Racial/Ethnic Groups | # Proposals | % Proposals | # PIs | % PIs | |
|---|---|---|---|---|---|
| **White, Not Hispanic** | 9,389 | 66% | 1,321 | 66% | |
| **Asian** | 3,880 | 27% | 537 | 27% | |
| **Hispanic/Latino** | 505 | 04% | 71 | 04% | **Included** |
| **Black/African American** | 245 | 02% | 50 | 02% | |
| Two or More Races | 130 | 01% | 17 | 01% | |
| Not Indicated | 106 | 01% | 17 | 01% | **Excluded** |
| American Indian/Alaskan Native | 8 | 00% | 1 | 00% | |
| TOTAL | 14,263 | 100% | 2,014 | 100% | |

Percentages may not sum to 100 due to rounding.

The number of proposals submitted, and the number of distinct PIs identified within each category are also shown.

It is often the case that the authors of unfunded applications are given the opportunity to modify the proposal based on reviewer feedback and resubmit. An earlier analysis from the NIH has shown that resubmissions are more likely to be funded than initial or "new" submissions [12]. Observations were, therefore, categorized based on whether proposals were new submissions or resubmissions according to the sponsor agency to which they were submitted. In the event that an observation was found to be a resubmission, only the final observation in the chain of submissions was retained for analysis. An example of this logic follows. Our example contains three observations: Proposal A, Proposal B, and Proposal C. If Proposal B is a resubmission of A, and Proposal C is a resubmission of B, then our final analysis discards Proposal A and Proposal B, keeping only Proposal C. Of the 14,263 proposals analyzed here, 2,489 (17%) are classified as resubmissions.

Although federal sponsors, such as the NIH, account for over 72% of total sponsored award dollars for the University of Michigan Medical School [13], the total population of individual sponsors across all extramural funding proposals is highly diverse with more than 1,320 unique sponsor organizations represented in our sample. Sponsors are, therefore, assigned to one of four general categories illustrated in Table 2 and based on the following criteria. Proposals to federal sponsors are divided into two subgroups. These proposals are designated as R01-equivalent if their NIH funding mechanism was R01 (n = 3,851) or DP1, DP2, DP5, R37, R56, RF1, RL1, U01, or R35 (n = 435). R01-equivalency is based on NIH categorization [14]. The remaining federal submissions were classified as Other Federal. Proposal submissions to industry were classified as Industry. Submissions to public charities, trade/professional organizations, foundations, international organizations, and other non-profit organizations were classified as Non-Profit. We label these synthetic groupings Submission Categories to differentiate them from the source feature, Sponsor Type. Proposal submissions to other sponsor types are excluded due to low sample size.

**Table 2. Assignment of UMMS Sponsor Types into Four Submission Categories (2010-2022).**

| UMMS Sponsor Type | # proposals | % | Submission Category | # proposals | % |
|---|---|---|---|---|---|
| **Federal** | 8,393 | 59% | **1: R01 & R01 equiv.** | 4,286 | 30% |
|  |  |  | **2. Other Federal** | 4,107 | 29% |
| **All Industry** | 1,102 | 08% | **3: Industry** | 1,102 | 08% |
| **International Organizations** | 30 | 00% | **4: Non-profit** | 4,745 | 33% |
| **All Trade/Professional Orgs** | 750 | 05% |  |  |  |
| **All Public Charities** | 2,263 | 16% |  |  |  |
| **All Foundations** | 1,015 | 07% |  |  |  |
| **All Other Non-Profit Orgs** | 687 | 05% |  |  |  |
| **All Foreign Government** | 8 | 00% | (excluded from our analysis due to insufficient sample size) | | |
| **US State & Local Authorities** | 15 | 00% |  |  |  |
| **TOTAL** | 14,263 |  |  | 14,240 |  |

Percentages may not sum to 100 due to rounding.
The data were analyzed using R version 4.1.3 for Windows (The R Foundation, Vienna, Austria).

## Indicators of observed award inequality and observed submission inequality

We created two indicators: award rate inequality (ARI), where proposals within the same submission categories are awarded at different rates according to the racial/ethnic group of the submitting PI and submission rate inequality (SRI), where the distribution of proposal submissions across submission categories is different according to the racial/ethnic group of the principal investigator.

ARI: The indicator is based on the ratio between the proposal award rate of the racial/ethnic group of interest and the proposal award rate of the majority racial/ethnic group, stratified by submission category. The majority racial/ethnic group is determined by number of unique PIs who self-identify with that group. Our indicator is based on the explicit assumption that the majority racial/ethnic group for each submission category will be positively correlated with a higher proposal award rate for that submission category. We therefore selected a measure that would allow for unambiguous comparison between each racial/ethnic group of interest and the majority. An ARI of 0.00 indicates award rate equity between the majority racial/ethnic group and the racial/ethnic group under consideration. A negative ARI value indicates a lower award rate for the group of interest relative to the majority while a positive ARI indicates a higher award rate for the group of interest relative to the majority. For example, assume that our data contains proposal award rates for two racial/ethnic groups, Group M and Group I. For the submission category being analyzed, 100 PIs identify with Group M. The award rate for Group M in this submission category is 0.45, meaning that 45% of Group M's proposal submissions in that submission category are awarded. Ten (10) PIs identify with Group I. The award rate for Group I in this same submission category is 0.39. Because Group I represents fewer total PIs than Group M in this submission category, we consider Group M our majority. Therefore, the ARI for Group I in this submission category would be calculated as: $0.39/0.45-1 = -0.13$.

SRI: The indicator is based on the ratio between the submission rate of the racial/ethnic group of interest and the submission rate of the majority racial/ethnic group, stratified again by submission category. When award rates between racial/ethnic groups are at parity then there is no a priori reason that submission rates would differ between these same groups.

# Results
## Expected impact of observed award rate on submission behavior

We presume that, in general, scientists can freely choose to which submission categories they submit their grant proposals (e.g., federal, industry, non-profit). While there are many factors that influence where scientists ultimately seek funding, there is no a priori constraint limiting PIs who identify with a certain racial/ethnic group from submitting a research grant proposal to any particular category of sponsor organization. We hypothesize, however, that perceived inequity in award rates between racial/ethnic groups when controlling for category of proposal submission will be correlated to differences in submission behavior. In other words, observed ARI, if lower within a particular submission category, acts as a constraint on sponsor choice, deterring PIs from submitting proposals to that category.

To further elaborate, PIs are presented with limited time in which to pursue their research. Each proposal consumes time that could otherwise be spent advancing their work elsewhere and is, therefore, viewed as an opportunity cost. To that end, PIs will seek to reduce the opportunity cost of proposal development by preferentially submitting grant proposals to submission categories where they have historically performed the best in terms of award rate and will have lower proposal submission rates to categories in which they have performed less well as a group.

## Observed award rates

As a first step we define the observed ARI of a given racial/ethnic group based on their performance relative to the majority racial/ethnic group. Throughout this paper we define the majority group for each submission category as that with the largest number of investigators in our sample. The most prevalent racial/ethnic group for awards across each submission category in our sample is White, which will therefore represent the 'baseline' for group comparison. The distribution of proposal awards by racial/ethnic group and submission category are displayed in Table 3.

**Table 3. Number of UMMS Awards by Racial/Ethnic Group and Submission Category (2010-2022).**

| Racial/Ethnic Group | R01/Equivalent | | Other Federal | | Industry | | Non Profit | | Total | |
|---|---|---|---|---|---|---|---|---|---|---|
| B/AA | 15 | 01% | 12 | 01% | 11 | 01% | 33 | 02% | 71 | 01% |
| Hispanic/Latino | 53 | 04% | 43 | 04% | 25 | 03% | 78 | 05% | 199 | 04% |
| Asian | 341 | 27% | 261 | 22% | 197 | 25% | 399 | 24% | 1,198 | 25% |
| White | 842 | 67% | 881 | 74% | 555 | 70% | 1,124 | 69% | 3,402 | 70% |
| TOTAL | 1,251 | | 1,197 | | 788 | | 1,634 | | 4,870 | |

$X^2$ (9, N = 4,870) = 21.46, p = .010. Percentages may not sum to 100 due to rounding.

The ratio of award rate by racial/ethnic group and submission category are illustrated in Fig 1. The expected impact on submission behavior for each group is derived in S1-S3 Tables. An exceptionally high rate of award for Industry submissions across all racial/ethnic categories was noted, which is not entirely unexpected as oftentimes industry will collaborate with PIs on areas of common interest. Industry may extend some opportunities directly to investigators based on specific expertise. Additionally, projects may be negotiated between investigators and industry partners informally prior to proposal development.

**Fig 1. Observed Award Rate.**
(A) Award Rate, illustrates the fraction of proposals awarded in each sponsor category by ethnic/racial group. The award rate relative to the majority group, shown in (B), is calculated in the following way:

$$A_r = \frac{A_{mina}/A_{mints}}{A_{maja}/A_{majts}} - 1$$

Where $A_r$ is the award ratio relative to the majority group, $A_{mina}$ is the total number of awards to the submission category by the given racial/ethnic group, $A_{mints}$ is the total number of submissions by the given racial/ethnic within the same submission category, $A_{maja}$ is the total number of awards to the sponsor category by the majority group, and $A_{majts}$ is the total number of submissions by the majority group within the same submission category. An example case of this calculation is given earlier.

We observed that both Asian and B/AA PIs at UMMS had lower award rates than White PIs across all submission categories at the group level, although the magnitude of inequality for each submission category differs between the two racial/ethnic groups. Asian PIs experience the highest award rates relative to White PIs in Industry submissions (-0.007), followed closely by R01 and Equivalent submissions (-0.053). Assuming these award rates translate to perceptions of bias by the PIs, as Hildreth [11] suggested, we predict that researchers who self-identify as Asian will consequently submit proposals to these submission categories at higher rates than to the Other Federal submission category, for instance, where award rate inequality vis-à-vis White PIs is greatest (-0.264).

B/AA researchers experience the highest award rates relative to the White researchers in the Non-Profit (-0.099) and Industry (-0.159) categories. As with Asian PIs, the award rates for B/AA PIs across all submission categories were negative relative to the White majority. In fact, B/AA researchers have the greatest negative ARI in every submission category for which there is sufficient data, except for Non-Profit, where they have a slightly higher ARI than their Asian colleagues (0.066). However, the magnitude of the negative ARI for B/AA PIs is significantly greater than that of Asian PIs.

Award rates for Hispanic/Latino PIs are higher than the majority in three of four categories: R01 and Equivalent (0.253), Industry (0.041), and Non-Profit (0.188). We would therefore expect this group of researchers to target these submission categories at higher rates than their White colleagues.

## Actual submission behaviors of each racial/ethnic group

Table 4 describes the distribution of proposal submissions across racial/ethnic groups and submission categories. The differences in proposal submission rates between groups are not random ($X^2$ (9, N = 13,996) = 16.39, p = .059); see S4-S6 Tables for proposal submission percentages and ratios by racial/ethnic group and submission category. As shown in Fig 2, each racial/ethnic group has a distinct submission pattern relative to White PIs. For instance, B/AA

PIs submit far fewer proposals to the R01 and Equivalent submission category than White PIs (-0.131) and far more of their proposals to Non-Profit sponsors (0.252).

**Table 4. Number of UMMS Submissions by Racial/Ethnic Group and Submission Category (2010-2022).**

| Racial/Ethnic Group | R01/Equivalent | | Other Federal | | Industry | | Non Profit | | Total | |
|---|---|---|---|---|---|---|---|---|---|---|
| B/AA | 64 | 02% | 62 | 02% | 18 | 02% | 100 | 02% | 244 | 02% |
| Hispanic/Latino | 142 | 03% | 151 | 04% | 33 | 03% | 179 | 04% | 505 | 04% |
| Asian | 1,209 | 29% | 1,092 | 27% | 273 | 25% | 1,303 | 28% | 3,877 | 28% |
| White | 2,827 | 67% | 2,714 | 68% | 763 | 70% | 3,066 | 66% | 9,370 | 67% |
| TOTAL | 4,242 | | 4,019 | | 1,087 | | 4,648 | | 13,996 | |

$X^2$ (9, N = 13,996) = 16.39, p = .059. Percentages may not sum to 100 due to rounding.

**Fig 2. Observed Submission Rate.**
(A) Submission Rate, illustrates the fraction of proposals submitted to each sponsor category by racial/ethnic group. The submission rate relative to the majority group, shown in (B), is calculated in the following way:

$$S_r = \frac{S_{mina}/S_{mints}}{S_{maja}/S_{majts}} - 1$$

Where $S_r$ is the submission ratio relative to the majority group, $S_{mina}$ is the total number of submissions to the sponsor category by the given racial/ethnic group, $S_{mints}$ is the total number of submissions by the given racial/ethnic group across all sponsor categories, $S_{maja}$ is the total number of submissions to the sponsor category by the majority group, and $S_{majts}$ is the total number of submissions by the majority group across all sponsor categories.

The rate of Industry submissions for each racial/ethnic group falls below the baseline for White PIs. Notable is the particularly low rate of Industry submissions by Hispanic/Latino PIs (-0.198), representing a reduction of 38% from the next lowest group, Asian (-0.135).

## Observed submission patterns of each racial/ethnic group

There is evidence of differential submission strategies by ethnic group ($X^2$ (9, N = 13,996) = 16.39, p = .059) at UMMS.

Submission and award ratios for the 12 possible category-group pairs (four submission categories and the three minority racial/ethnic groups) are only weakly associated (see S7 Table, which shows correspondence between submission and award ratios by racial/ethnic group). Notably, award rates for White investigators exceeded those of every other racial/ethnic group in nine of 12 (75%) submission categories. Although Hispanic/Latino investigators had higher rates of award than the majority in R01/Equivalent (ARI: 0.253), Industry (ARI: 0.041) and Non-Profit (ARI: 0.188) submission categories, submission rates within these categories for this group did not correspondingly increase as was expected.

## Discussion

This study identified racial/ethnic differences in proposal submission and award rates from external sponsors for PIs at the University of Michigan Medical School. To the best of our knowledge, this is the first time an analysis of this kind has been conducted on a comprehensive set of proposals by a well-defined population of researchers. Despite evidence that proposal submission patterns differ between racial/ethnic groups, we did not find a significant correlation between submission strategies and award rates. However, there is a clear relationship between investigator race/ethnicity and research proposal award rates.

Our findings are consistent with prior reports of lower NIH award rates to B/AA investigators: B/AA PIs within the UMMS are not submitting R01 and equivalent proposals as often as their peers and when they do submit, their proposals are awarded less frequently. Indeed, the award rates for B/AA PIs lag behind other racial/ethnic groups in every submission category except for Industry, where B/AA award rates are slightly higher than the Asian group. Of note, award rates for B/AA PIs fell furthest behind their White colleagues in the submission categories of R01/Equivalent and Other Federal, suggesting that previous findings of B/AA funding disadvantage within the NIH may extend to other federal biomedical research funding.

These findings were not isolated to B/AA PIs. Our results show that both Asian and B/AA PIs at the UMMS experience lower award rates than their White peers across all sponsor submission categories. Due to relatively high representation within biomedicine, Asian PIs are not considered an under-represented minority. The unexpected finding that Asian PIs appear disadvantaged relative to their White peers across all submission categories indicates that funding inequality is not simply a function of population.

While the ARI for both R01/Equivalent and Other Federal submission categories were lowest for B/AA PIs followed by Asian PIs (B/AA: -0.213, -0.403; Asian: -0.053, -0.264), the ARI ratio between these submission categories is particularly noteworthy. ARI for Other Federal submissions was 62% worse than R01/Equivalent submissions for B/AA PIs and 133% worse for Asian PIs. The Other Federal category represents everything from large Program Project/Center Grants (P series) to Phase 1 exploratory mechanisms within NIH, and all awards from other federal agencies such as the National Science Foundation and the Department of Defense.

It is important to restate that our observations are limited to a single institution; therefore, we cannot make generalizable claims about either sponsors or racial/ethnic groups, and thus, further investigation at other institutions is warranted to address whether there are systemic disadvantages among the various submission categories for racial/ethnic groups, particularly B/AA PIs. Nonetheless, we believe these results signal an urgent need for other academic institutions to replicate our results, and for other federal agencies to replicate the analyses of the NIH on their own proposal data to better understand these differentials.

Hoppe, *et al.* found that there may be topic biases in funding institutions that effectively constrain submission category choice [5]. A more recent publication by Lauer indicated that the funding disparity for B/AA PIs was not a product of the peer reviewers, but a result of their applications being assigned to NIH Institutes and Centers with lower award rates [6]. Lower funding rates are not the only issue that researchers may face when deciding where to submit. Certain topics may only be funded by relatively few sponsor organizations, or only by sponsors of a certain category, in which case pursuing the given topic would reduce the available funding opportunity significantly. Hoppe's previously mentioned study also touched on this point. Further investigations into topic bias and availability of funding by topic across different racial/ethnic groups would be a fruitful direction for future research.

Due to small sample sizes when stratifying proposal observations by both race/ethnicity of PI and proposal/award submission category, we were unable to control for some characteristics of PI submission behavior that have been previously shown to impact award rates. For instance, Doyle, *et al.* have found that first-time R01 applicants who resubmitted their original unfunded proposals were more likely to receive R01 funding within 3 and 5 years than applications who did not [16]. Haggerty and Fenton found that submission behaviors such as submitting proposals to more than one NIH Institute, submitting more applications per year, and ability to write proposals that were scored vs triaged were all correlated with greater likelihood of NIH award [7]. In addition, they found that there were certain demographic characteristics of PIs—such as possession of an MD or MD/PhD—that were correlated with NIH funding success.

These limitations represent a well-defined path for future study but will require access to a larger pool of proposal data. A shortcoming of the current literature is that analysis is primarily sponsor-specific: the proposal portfolio of a single sponsor agency is analyzed. This study sought to analyze research proposals across all sponsors within a single academic medical center; however, this approach limited the pool of proposals available to study and resulted in small sample size for some groups. Another notable consequence of this was the insufficient representation of PIs identifying with certain racial/ethnic sub-demographics for statistical analysis, such as American Indian/Alaskan Native or Native Hawaiian/Pacific Islander.

An opportunity for greater potential future impact—including better representation across racial/ethnic groups—will be to coordinate the analysis of research proposals across multiple academic institutions. Considering this opportunity, we note that neither personally identifiable data about PIs nor "sensitive" proposal data—such as specific aims or the research strategies of unfunded research proposals—are required for this analysis. So long as institutions agree on a relatively simple query design and data aggregation methodology, they may share the proposal and PI demographic data required to replicate this study without exposing confidential information about their PI cohort or the intellectual property contained within their proposals. An interinstitutional analysis of this type would represent a significant contribution.

The favorable award rates for Hispanic/Latinos also warrants further inquiry and would benefit from interinstitutional analysis, testing whether the observation is differential to UMMS or more generalizable. Given the positive award rates observed across multiple sponsor categories, it would be interesting to determine the underlying causes and whether particular practices or strategies employed by these investigators could be applied by others.

The analysis of research proposal databases is a critical step in understanding how academic institutions can better support researchers and how sponsor agencies can better position themselves to reduce research funding inequity. Introspective analysis of an institution's grant portfolio may reveal disparities in funding patterns like those found here and, more importantly, illuminate opportunities for positive institutional intervention. Interventions can take many forms. Committees may de-emphasize the role of R01 and equivalent awards in promotion and tenure review, de-emphasize the communication of external rankings with known biases, and emphasize non-financial metrics—such as participation in diverse research teams, invisible work, and mentorship—in evaluative exercises. Support could also be provided during the grant writing process, with training and mentorship on this topic as early as the doctorate level.

Additionally, data can spark much-needed conversations about institutional blind spots that risk perpetuating systemic inequities. As a scientific community, we are guided by data. As universities, funders, and individual researchers commit to building a more inclusive research community, the analysis of institutional proposal databases offers a compelling opportunity for novel insight that allows for communal self-reflection, and a guide for future improvements.

# Acknowledgments
We thank the reviewers for their critical reading and insight. Their criticism improved our analysis and this manuscript.

# Funding/Support
None

# Other Disclosures
None

# Disclaimers

None

# Previous Presentations

None

# Supporting information

**S1 Table. Proposal Award Rates for Proposals from Asian vs. White Researchers.**

**S2 Table. Proposal Award Rates for Proposals from B/AA vs White Researchers.**

**S3 Table. Proposal Award Rates for Proposals from Hispanic/Latino vs. White Researchers.**

**S4 Table. Proposal Submission Percentages by Asian vs. White Researchers.**

**S5 Table. Proposal Submission Percentages by B/AA vs. White Researchers.**

**S6 Table. Proposal Submission Percentages by Hispanic/Latino vs. White Researchers.**

**S7 Table. Comparison of Submission and Award Rates.** The award and submission ratios for each racial/ethnic group-submission category pair.

**S8 File. Proposal Submission and Award Data by Submission Category and PI Race/Ethnicity.**

Figure 1

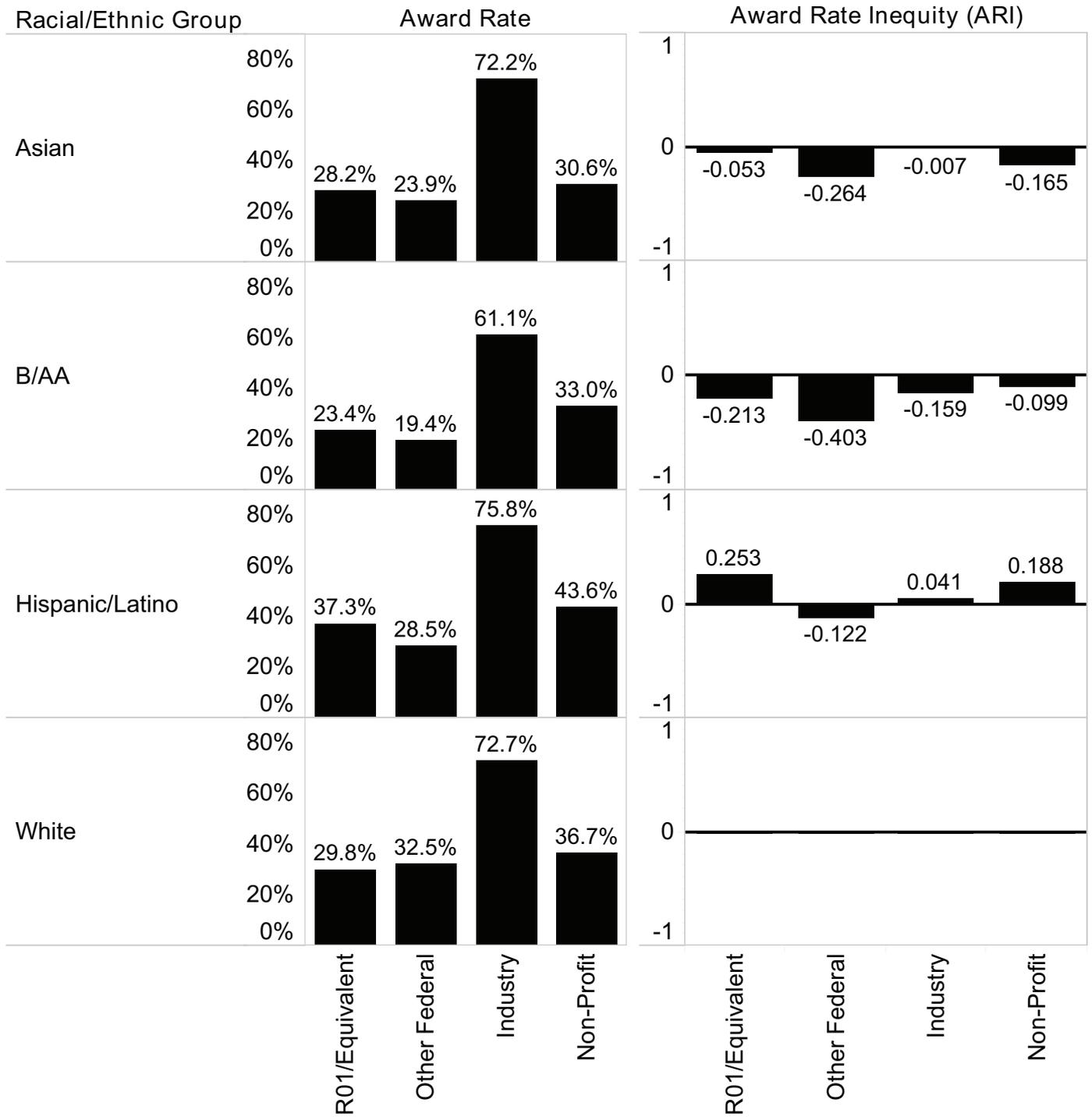

Figure 2

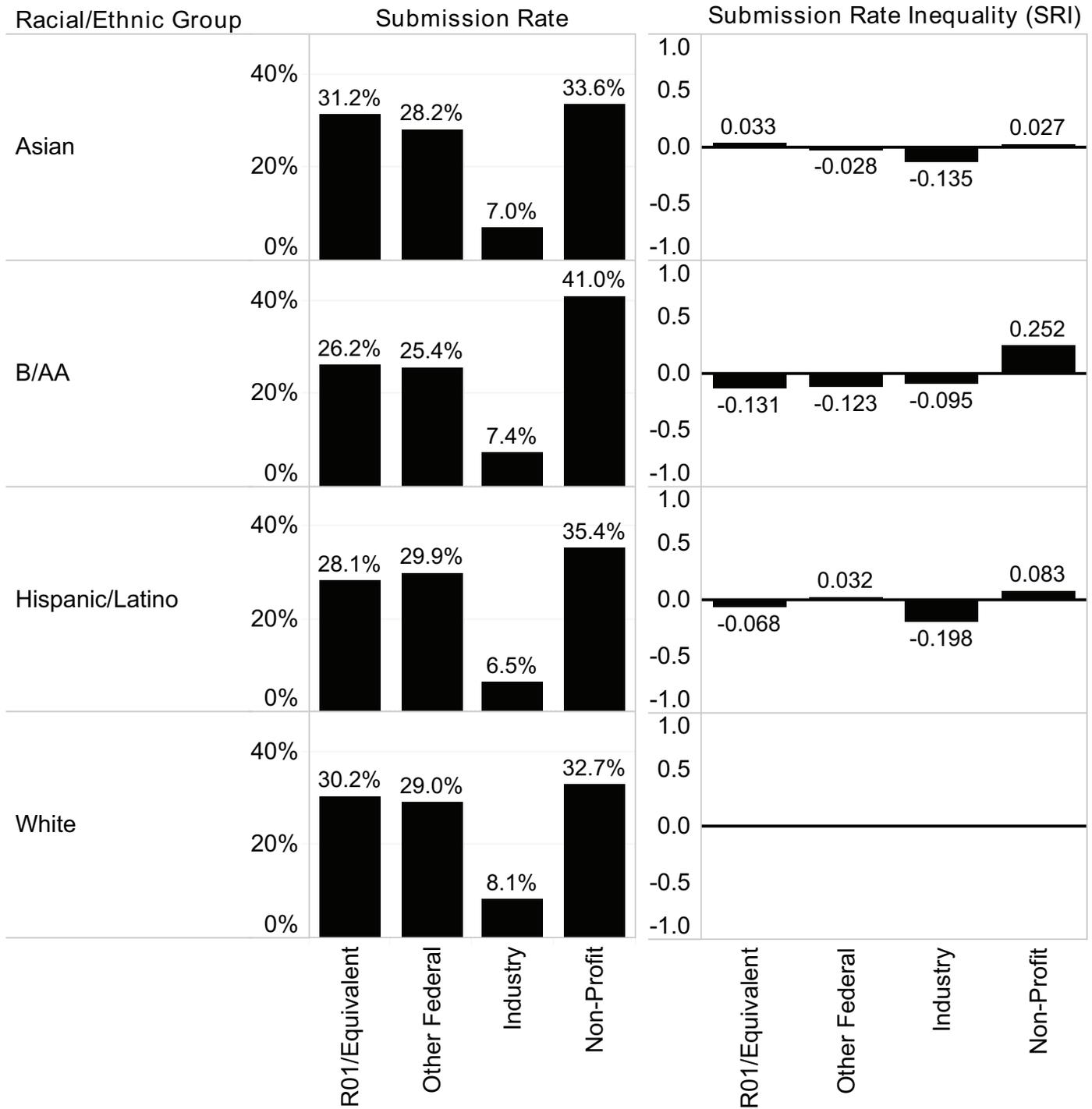

S1 TABLE

|  | R01/Equivalent | Other Federal | Industry | Non-Profit |
|---|---|---|---|---|
| Asian | 28.20% | 23.90% | 72.16% | 30.62% |
| White | 29.78% | 32.46% | 72.73% | 36.66% |
| Ratio | -0.053 | -0.264 | -0.007 | -0.165 |
| Impact | Negative | Negative | Negative | Negative |

S2 TABLE

|       | R01/Equivalent | Other Federal | Industry | Non-Profit |
|-------|---------------|---------------|----------|------------|
| B/AA  | 23.43%        | 19.35%        | 61.11%   | 33.00%     |
| White | 29.78%        | 32.46%        | 72.73%   | 36.66%     |
| Ratio | -0.213        | -0.403        | -0.159   | -0.099     |
| Impact| Negative      | Negative      | Negative | Negative   |

S3 TABLE

|  | R01/Equivalent | Other Federal | Industry | Non-Profit |
|---|---|---|---|---|
| Hispanic/Latino | 37.32% | 28.47% | 75.75% | 43.57% |
| White | 29.78% | 32.46% | 72.73% | 36.66% |
| Ratio | 0.253 | -0.122 | 0.041 | 0.188 |
| Impact | Positive | Negative | Positive | Positive |

S4 TABLE

|       | R01/Equivalent | Other Federal | Industry | Non-Profit |
|-------|----------------|---------------|----------|------------|
| Asian | 31.18%         | 28.16%        | 7.04%    | 33.60%     |
| White | 30.17%         | 28.96%        | 8.14%    | 32.72%     |
| Ratio | 0.033          | -0.028        | -0.135   | 0.027      |

S5 TABLE

|  | R01/Equivalent | Other Federal | Industry | Non-Profit |
|---|---|---|---|---|
| B/AA | 26.22% | 25.40% | 7.37% | 40.98% |
| White | 30.17% | 28.96% | 8.14% | 32.72% |
| Ratio | -0.131 | -0.123 | -0.095 | 0.252 |

S6 TABLE

|  | R01/Equivalent | Other Federal | Industry | Non-Profit |
|---|---|---|---|---|
| Hispanic/Latino | 28.11% | 29.90% | 6.53% | 35.44% |
| White | 30.17% | 28.96% | 8.14% | 32.72% |
| Ratio | -0.068 | 0.032 | -0.198 | 0.083 |

S7 TABLE

|  | R01 /Equivalent | Other Federal | Industry | Non-Profit |
|---|---|---|---|---|
| **Asian** | + - | - - | - - | + - |
| **B/AA** | - - | - - | - - | + - |
| **Hispanic/Latino** | - + | + - | - + | + + |

| | |
|---|---|
| + + | Submission and Award ratios are both positive relative to majority |
| + - | Submission ratio is positive and Award ratio is negative relative to majority |
| - - | Submission and Award ratios are both negative relative to majority |
| - + | Submission ratio is negative and Award ratio is positive relative to majority |